\documentclass[12pt]{article}
\usepackage{graphicx}
\usepackage{cite}
\newcommand{\ind}[2]{^{#1}_{\mbox{\scriptsize #2}}}
\newcommand{\inds}[2]{^{#1}_{\mbox{\tiny #2}}}
\newcommand{\al}[2]{\alpha\ind{#1}{#2}}
\newcommand{\ASL}[2]{A^{#1}_{\mbox{\tiny SL},\,#2}}
\newcommand{\ATL}[2]{A^{#1}_{\mbox{\tiny TL},\,#2}}
\newcommand{\ro}[2]{\varrho^{#1}_{#2}}
\newcommand{\ARe}[1]{A\ind{#1}{Re}}
\newcommand{\AIm}[1]{A\ind{#1}{Im}}
\newcommand{\Sy}[2]{S\ind{(#1)}{#2}(y)}
\newcommand{\subheader}[1]{\noindent\vspace*{12.5mm}\raisebox{0.8pt}{$\bullet\:$}\texttt{#1}\par\vspace*{-12.5mm}}
\newcommand{\cmd}[1]{\texttt{\underline{#1}}$\,$}
\def\nf{n_{\mbox{\scriptsize f}}}
\def\MSbar{$\overline{\mbox{MS}}$}

\begin{document}

\begin{titlepage}
\begin{center}

{\Large\bf QCDMAPT: program package for \\ Analytic approach to QCD

}

\vskip10mm

{\large A.V.~Nesterenko$^{a,}$\footnote[1]{E-mail: nesterav@theor.jinr.ru}
and C.~Simolo$^{b}$}

\vskip5mm

$^{a}${\small\it BLTPh, Joint Institute for Nuclear Research,
Dubna, 141980, Russian Federation}

\vskip1mm

$^{b}${\small\it ISAC--CNR, I--40129, Bologna, Italy}

\end{center}

\vskip5mm
\hrule
\vskip2.5mm

\noindent
{\bf Abstract}

\vskip2.5mm

\noindent
A program package, which facilitates computations in the framework of
Analytic approach to QCD, is developed and described in details. The
package includes the explicit expressions for relevant spectral functions
calculated up to the four--loop level and the subroutines for necessary
integrals.

\vskip5mm


\vskip2.5mm


\vskip2.5mm
\hrule
\vskip7.5mm

\noindent
{\bf PROGRAM SUMMARY}
\vskip2.5mm

\begin{small}
\noindent
{\em Manuscript Title:} QCDMAPT: program package for
     Analytic approach to QCD                                 \\
{\em Authors:} A.V.~Nesterenko and C.~Simolo                  \\
{\em Program Title:} QCDMAPT                                   \\
{\em Journal Reference:}                                      \\
{\em Catalogue identifier:}                                   \\
{\em Licensing provisions:} none                              \\
{\em Programming language:} Maple~9 and higher                \\
{\em Computer:} Any which supports Maple~9                    \\
{\em Operating system:} Any which supports Maple~9            
{\em Classification:} 11.1, 11.5, 11.6 \\
%
{\em Nature of problem:} Subroutines helping computations within
     Analytic approach to~QCD \\
{\em Solution method:} A program package for Maple is provided.
     It includes the explicit expressions for relevant spectral
     functions and the subroutines for basic integrals used in
     the framework of Analytic approach to QCD. \\
{\em Running time:} Template program running time is about a minute
     (depends on~CPU)
\end{small}

\end{titlepage}

\newpage

\setcounter{page}{2}

\section{Introduction}

The strong interactions display two fundamental features, namely, the
asymptotic freedom (at high energies) and color confinement (at low
energies). The first feature allows one to study the strong interaction
processes in the ultraviolet domain by making use of perturbation theory.
However, theoretical description of hadron dynamics at low energies
requires nonperturbative methods.

In general, there is a variety of nonperturbative approaches to handle the
strong interaction processes at low energies. In this work we will focus
on the so--called ``dispersive'' (or ``analytic'') approach to Quantum
Chromodynamics~(QCD). The basic idea of this approach is to merge the
perturbative results with the nonperturbative constraints arising from
relevant dispersion relations. In turn, this eliminates some intrinsic
difficulties of perturbation theory and extends its range of applicability
towards the infrared domain. One of the possible implementations of this
approach within QCD is the so--called ``massless'' Analytic Perturbation
Theory~\cite{APT1,APT2,APT3}. The latter has been successfully employed in
the studies of various strong interaction processes (see, e.g.,
papers~\cite{Prosperi1,Prosperi2,Prosperi3,PRD,CSB,SR,PionFF,BMS,Kotikov,Arbuzov,Cvetic},
reviews~\cite{APT3,Prosperi4,Review} and references therein). The
incorporation of effects due to the nonvanishing mass of lightest hadron
state has been implemented into approach in hand within the so--called
``massive'' Analytic Perturbation Theory, see Refs.~\cite{MAPT1,MAPT2} for
the details.

A central object of the current approach is the so--called spectral
function, which can be calculated by making use of the strong running
coupling~$\al{}{s}(Q^2)$ (see Sect.~\ref{Sect:RhoCalc}). At the one--loop
level the perturbative running coupling and relevant spectral function
have a quite simple form (Eqs.~(\ref{aPert1L}) and~(\ref{Rho1L1p}),
respectively). However, at the higher loop levels the strong running
coupling has a rather cumbersome structure, and the calculation of
corresponding spectral functions\footnote{At the higher loop levels the
spectral functions can also be calculated numerically. However, it
requires a lot of computation resources and essentially slows down the
overall computation process.} represents a rather complicated task.

The primary objective of this paper is to calculate (by~hands) the
explicit expressions for the aforementioned spectral functions at the
higher loop levels and to incorporate them (together with proper
subroutines for necessary integrals) into a single program package. In
turn, the latter will facilitate computations within the approach in hand.

The layout of the paper is as follows. Section~\ref{Sect:APT} constitutes
a brief description of the Analytic approach to~QCD. The calculation of
spectral functions and the numerical evaluation of basic integrals is
considered in Sects.~\ref{Sect:RhoCalc} and~\ref{Sect:Int}, respectively.
Section~\ref{Sect:Template} describes the template program and includes
the list of package commands. The basic results are summarized in the
Conclusions (Sect.~\ref{Sect:Concl}). Appendix~\ref{Sect:RCPert} contains
explicit expressions for the perturbative strong running coupling up to
the four--loop level. Appendix~\ref{Sect:Rho} contains explicit
expressions for the spectral functions calculated up to the four--loop
level.

\section{Analytic approach to QCD}
\label{Sect:APT}

In general, in the framework of perturbation theory the high energy
behavior of the strong correction~$d(Q^2)$ to a physical
observable~$D(Q^2)$ can be approximated by the power series in the strong
running coupling~$\al{}{s}(Q^2)$. Namely, at the $\ell$--loop level
\begin{equation}
\label{dPert}
d\ind{(\ell)}{pert}(Q^2) =
\sum_{j=1}^{\ell} d_{j} \Bigl[\al{(\ell)}{s}(Q^2)\Bigr]^{j} =
\sum_{j=1}^{\ell} d_{j} \biggl(\frac{4\pi}{\beta_0}\biggr)^{\! j}
\Bigl[a\ind{(\ell)}{s}(Q^2)\Bigr]^{j}, \qquad Q^2 \to \infty,
\end{equation}
where $Q^2 = -q^2 > 0$ is the spacelike kinematic variable, $d_{j}$ stands
for the relevant perturbative expansion coefficient,
$\al{(\ell)}{s}(Q^2)$~is the $\ell$--loop perturbative strong running
coupling (see App.~\ref{Sect:RCPert}), $\beta_0 = 11 - 2 \nf/3$ denotes
the one--loop perturbative $\beta$~function expansion coefficient,
$\nf$~is the number of active quarks, and $a\ind{(\ell)}{s}(Q^2) \equiv
\al{(\ell)}{s}(Q^2) \beta_{0}/(4\pi)$ is the so--called ``couplant''.
However, perturbative expansion~(\ref{dPert}) is valid in the ultraviolet
domain only. In particular, at any loop level
$d\ind{(\ell)}{pert}(Q^2)$~(\ref{dPert}) possesses unphysical
singularities in the infrared domain. This fact contradicts the general
principles of local Quantum Field Theory and essentially complicates the
analysis of low~energy experimental data. Besides, perturbative
expansion~(\ref{dPert}) can not be directly employed in the theoretical
description of physical observables depending on the timelike\footnote{For
example, the experimental data on the so--called $R(s)$--ratio of
electron--positron annihilation into hadrons can only be examined by
making use of both perturbation theory and relevant dispersion relation,
see, e.g., Ref.~\cite{Adler}.} kinematic variable~$s=q^2>0$.

As it has been noted in the Introduction, one can overcome the
aforementioned difficulties of perturbative approach by invoking relevant
dispersion relations. Thus, in the framework of ``massless'' Analytic
Perturbation Theory (APT)~\cite{APT1,APT2,APT3} the theoretical
expressions for the strong corrections~$d(Q^2)$ and~$r(s)$ to physical
observables~$D(Q^2)$ and~$R(s)$, depending on spacelike ($Q^2=-q^2>0$) and
timelike ($s=q^2>0$) kinematic variables, take the form (see
papers~\cite{APT2,APT3} and references therein for the details)
\begin{equation}
\label{dAPT}
d\inds{(\ell)}{APT}(Q^2) =
\sum_{j=1}^{\ell} d_{j} \biggl(\frac{4\pi}{\beta_0}\biggr)^{\! j}
\bar\ASL{(\ell)}{j}(z), \quad
\bar\ASL{(\ell)}{j}(z) = \int\limits_{0}^{\infty}
\frac{\ro{(\ell)}{j}(\sigma)}{\sigma + z} \, d \sigma, \quad
z = \frac{Q^2}{\Lambda^2},
\end{equation}
\begin{equation}
\label{rAPT}
r\inds{(\ell)}{APT}(s) =
\sum_{j=1}^{\ell} d_{j} \biggl(\frac{4\pi}{\beta_0}\biggr)^{\! j}
\bar\ATL{(\ell)}{j}(w), \quad
\bar\ATL{(\ell)}{j}(w) = \int\limits_{w}^{\infty}
\ro{(\ell)}{j}(\sigma)\,\frac{d \sigma}{\sigma}, \quad
w = \frac{s}{\Lambda^2}.
\end{equation}
Here and further $\Lambda$ denotes the QCD scale parameter and
$\ro{(\ell)}{j}(\sigma)$ stands for the spectral function corresponding to
$j$-th power of the $\ell$--loop couplant (see Sect.~\ref{Sect:RhoCalc}).

In general, the effects due to the mass of the lightest hadron state can
be safely neglected at intermediate and high energies only. In particular,
such effects play an essential role in theoretical description of the
strong interaction processes at low energies. As it has been noted in the
Introduction, the effects due to the nonvanishing mass of the lightest
hadron state have been accounted for within so--called ``massive''
Analytic Perturbation Theory (MAPT)~\cite{MAPT1,MAPT2}. Thus, in the
framework of MAPT the theoretical expressions for the above--mentioned
strong corrections read
\begin{equation}
\label{dMAPT}
d\inds{(\ell)}{MAPT}(Q^2,m^2) =
\sum_{j=1}^{\ell} d_{j} \biggl(\frac{4\pi}{\beta_0}\biggr)^{\! j}
\ASL{(\ell)}{j}(z,\chi), \qquad z = \frac{Q^2}{\Lambda^2}, \quad
\chi = \frac{m^2}{\Lambda^2},
\end{equation}
\begin{equation}
\label{ASLMAPT}
\ASL{(\ell)}{j}(z,\chi) = \frac{z}{z+\chi}
\int\limits_{\chi}^{\infty} \ro{(\ell)}{j}(\sigma)\,
\frac{\sigma - \chi}{\sigma+z}\, \frac{d \sigma}{\sigma},
\end{equation}
\begin{equation}
\label{rMAPT}
r\inds{(\ell)}{MAPT}(s,m^2) =
\sum_{j=1}^{\ell} d_{j} \biggl(\frac{4\pi}{\beta_0}\biggr)^{\! j}
\ATL{(\ell)}{j}(w,\chi), \qquad w = \frac{s}{\Lambda^2},
\end{equation}
\begin{equation}
\label{ATLMAPT}
\ATL{(\ell)}{j}(w,\chi) =
\theta(w-\chi)\int\limits_{w}^{\infty}
\ro{(\ell)}{j}(\sigma)\,\frac{d \sigma}{\sigma},
\end{equation}
where $\theta(x)$ is the unit step function ($\theta(x)=1$ if $x \ge 0$
and $\theta(x)=0$ otherwise). It is worth noting that in the limit~$m \to
0$ Eqs.~(\ref{dMAPT}) and~(\ref{rMAPT}) coincide with Eqs.~(\ref{dAPT})
and~(\ref{rAPT}), respectively:
$d\inds{(\ell)}{MAPT}(Q^2,0)=d\inds{(\ell)}{APT}(Q^2)$ and
$r\inds{(\ell)}{MAPT}(s,0)=r\inds{(\ell)}{APT}(s)$. Besides,
$d\inds{(\ell)}{MAPT}(Q^2,m^2) \to d\inds{(\ell)}{APT}(Q^2)$ for $Q^2 \gg
m^2$ and $r\inds{(\ell)}{MAPT}(s,m^2)=r\inds{(\ell)}{APT}(s)$ for $s>m^2$
(see Refs.~\cite{MAPT1,MAPT2} for the details). It is worthwhile to
emphasize also that expressions (\ref{dAPT}), (\ref{rAPT}) and
(\ref{dMAPT})--(\ref{ATLMAPT}) represent the nonpower functional
expansions\footnote{In other words, for $j \ge 2$ $\bar\ASL{(\ell)}{j}(z)
\neq \Bigl[\bar\ASL{(\ell)}{1}(z)\Bigr]^{j}$ and $\bar\ATL{(\ell)}{j}(w)
\neq \Bigl[\bar\ATL{(\ell)}{1}(w)\Bigr]^{j}$. Nonetheless, in the
ultraviolet asymptotic ($z\to\infty$) $\bar\ASL{(\ell)}{j}(z) \to
\bigl[a\ind{(\ell)}{s}(Q^2)\bigr]^{j}$, i.e., the expansions~(\ref{dAPT})
and~(\ref{dMAPT}) reproduce the perturbative power series~(\ref{dPert}).}
of the strong corrections~$d(Q^2)$ and~$r(s)$.

\section{Calculation of the spectral functions}
\label{Sect:RhoCalc}

\begin{figure}[t]
\centerline{\includegraphics[width=110mm,clip]{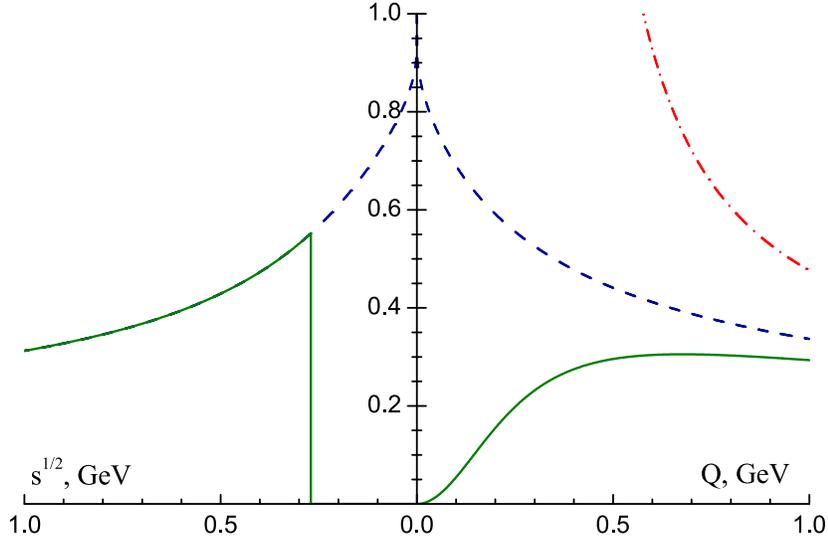}}
\caption{The one--loop perturbative couplant (Eq.~(\ref{aPert1L}),
dot--dashed curve) and the one--loop first--order expansion
functions: APT (Eq.~(\ref{APT1L}), dashed curves) and MAPT
(Eqs.~(\ref{MAPTSL1L}),~(\ref{MAPTTL1L}), solid curves).
The values of parameters: $\Lambda=350\,$MeV, $m=270\,$MeV.}
\label{Plot:A1L}
\end{figure}

In general, there is no unique way to restore the aforementioned spectral
function~$\ro{(\ell)}{j}(\sigma)$ by making use of the perturbative
expression for the strong running coupling~$\al{(\ell)}{s}(Q^2)$
(discussion of this issue can be found in Refs.~\cite{PRD,Review,MPLA2}).
In what follows we adopt the definition proposed in
Refs.~\cite{APT1,APT2,APT3}:
\begin{equation}
\label{RhoDef}
\ro{(\ell)}{j}(\sigma) = \frac{1}{2 \pi i} \lim_{\varepsilon \to 0_{+}}
\Biggl(\biggl\{a\ind{(\ell)}{s}\Bigl[-\Lambda^2(\sigma + i \varepsilon)\Bigr]\!\biggr\}^{j} -
\biggl\{a\ind{(\ell)}{s}\Bigl[-\Lambda^2(\sigma - i \varepsilon)\Bigr]\!\biggr\}^{\!j}\Biggr)
\end{equation}
($\sigma$ is a dimensionless variable). At the one--loop level~($\ell=1$)
the first--order~(\mbox{$j=1$}) spectral function~(\ref{RhoDef}) can
easily be calculated (see Eq.~(\ref{aPert1L})):
\begin{eqnarray}
\ro{(1)}{1}(\sigma) &=& \frac{1}{2 \pi i} \lim_{\varepsilon \to 0_{+}}
\biggl(a\ind{(1)}{s}\Bigl[-\Lambda^2(\sigma + i \varepsilon)\Bigr] -
a\ind{(1)}{s}\Bigl[-\Lambda^2(\sigma - i \varepsilon)\Bigr]\biggr)
\nonumber \\[1.5mm]
\label{Rho1L1p}
&=& \frac{1}{2 \pi i} \lim_{\varepsilon \to 0_{+}}
\Biggl[\frac{1}{\ln(-\sigma - i \varepsilon)} -
\frac{1}{\ln(-\sigma + i \varepsilon)}\Biggr] =
\frac{1}{y^2 + \pi^2},
\end{eqnarray}
where $y=\ln\sigma$. Eventually, this leads to the following
expressions\footnote{It is assumed that $\arctan x$ is a continuously
increasing function of its argument: $-\pi/2\leq\arctan x\leq\pi/2$ for
$-\infty < x < \infty$.} for the one--loop ($\ell=1$) first--order ($j=1$)
expansion functions~(\ref{dAPT}), (\ref{rAPT}), (\ref{ASLMAPT}),
and~(\ref{ATLMAPT}):
\begin{equation}
\label{APT1L}
\bar\ASL{(1)}{1}(z) = \frac{1}{\ln z} + \frac{1}{1-z}, \qquad \quad
\bar\ATL{(1)}{1}(w) =
\frac{1}{2} - \frac{1}{\pi}\arctan\!\left(\frac{\ln w}{\pi}\right)\!,
\end{equation}
\begin{equation}
\label{MAPTSL1L}
\ASL{(1)}{1}(z,\chi) = \frac{1}{\ln z} +
\frac{z}{1-z}\,\frac{1+\chi}{z+\chi} -
\frac{z}{z+\chi}\int\limits_{0}^{\chi}
\frac{\ro{(1)}{1}(\sigma)}{\sigma+z}
\biggl(1-\frac{\chi}{\sigma}\biggr)\, d \sigma,
\end{equation}
\begin{equation}
\label{MAPTTL1L}
\ATL{(1)}{1}(w,\chi) = \theta(w-\chi) \Biggl[\,
\frac{1}{2} - \frac{1}{\pi}\arctan\!\left(\frac{\ln w}{\pi}\right)\Biggr],
\end{equation}
see papers~\cite{APT3,MAPT2} and references therein for the details. The
functions (\ref{APT1L})--(\ref{MAPTTL1L}) are shown in
Fig.~\ref{Plot:A1L}.

At the higher loop levels ($\ell \ge 2$) perturbative couplants
$a\ind{(\ell)}{s}(Q^2)$ have a cumbersome structure (see
App.~\ref{Sect:RCPert}). Hence, the calculation of the $\ell$--loop
spectral function of $j$--th order~$\ro{(\ell)}{j}(\sigma)$~(\ref{RhoDef})
($1 \le j \le \ell$) represents a rather complicated task. The numerical
calculation of the spectral functions~(\ref{RhoDef}) requires a lot of
computational resources and essentially slows down the running of the
program. Nonetheless, this problem can be resolved\footnote{An alternative
way to overcome this problem is to construct a set of explicit expressions
which approximate the nonpower expansion functions~(\ref{dAPT}),
(\ref{rAPT}), (\ref{ASLMAPT}), and~(\ref{ATLMAPT}) accurately enough, see,
e.g., Ref.~\cite{Approx}.} in the following way.

\begin{figure}[t]
\centerline{\includegraphics[width=110mm,clip]{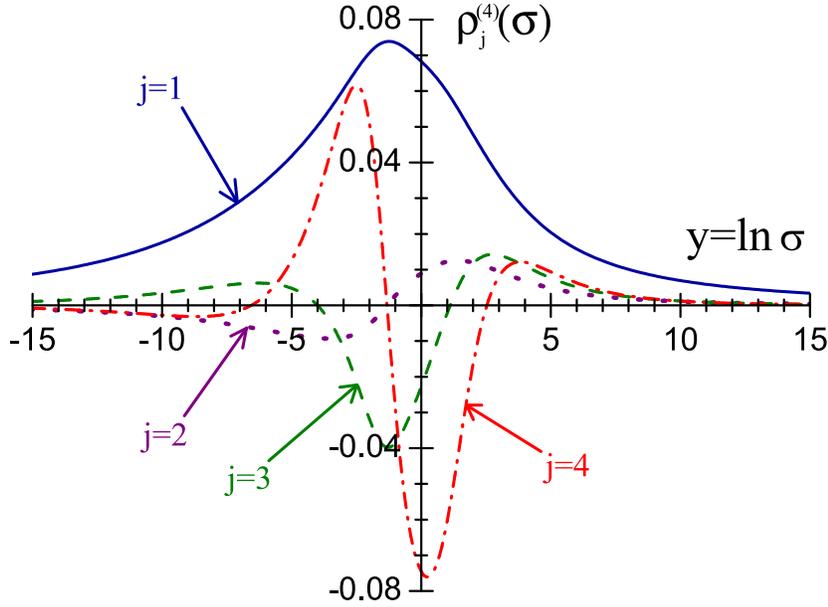}}
\caption{The four--loop spectral functions
$\ro{(4)}{j}(\sigma)$~(\ref{RhoDef}): $\ro{(4)}{1}(\sigma)$~(solid curve),
$\ro{(4)}{2}(\sigma)$~(dotted curve), $10\cdot\ro{(4)}{3}(\sigma)$~(dashed
curve), and $10^{2}\cdot\ro{(4)}{4}(\sigma)$~(dot--dashed curve). The
values of parameters: $\nf=3$, $y=\ln\sigma$.}
\label{Plot:Rho}
\end{figure}

First of all, it is convenient to express the spectral functions
$\ro{(\ell)}{j}(\sigma)$~(\ref{RhoDef}) in terms of the real and imaginary
parts of the $\ell$--loop couplant~$a\ind{(\ell)}{s}(Q^2)$ on a physical
cut:
\begin{equation}
\label{AReImDef}
\lim_{\varepsilon \to 0_{+}}
a\ind{(\ell)}{s}\Bigl[-\Lambda^2(\sigma \mp i\varepsilon)\Bigr] \equiv
\ARe{(\ell)}(\sigma) \mp i \pi \AIm{(\ell)}(\sigma).
\end{equation}
In this case Eq.~(\ref{RhoDef}) can be represented in a concise form:
\begin{equation}
\label{RhoPartHO}
\ro{(\ell)}{j}(\sigma) = \sum_{k=0}^{K_j}
{j \choose 2k+1}
(-1)^{k}\, \pi^{2k}
\Bigl[\AIm{(\ell)}(\sigma)\Bigr]^{2k+1}
\Bigl[\ARe{(\ell)}(\sigma)\Bigr]^{j-2k-1},
\end{equation}
where $K_j = j/2 + (j \;\mbox{mod}\; 2)/2 -1$, and
\begin{equation}
{n \choose m} = \frac{n!}{m!\,(n-m)!}
\end{equation}
is the binomial coefficient. In particular, at any loop level
\begin{eqnarray}
\ro{(\ell)}{1}(\sigma) &=& \AIm{(\ell)}(\sigma), \\[2.5mm]
\ro{(\ell)}{2}(\sigma) &=& 2 \AIm{(\ell)}(\sigma)
     \ARe{(\ell)}(\sigma), \\[2.5mm]
\ro{(\ell)}{3}(\sigma) &=& \AIm{(\ell)}(\sigma)
     \Biggl\{3\Bigl[\ARe{(\ell)}(\sigma)\Bigr]^{2} -
     \pi^{2}\Bigl[\AIm{(\ell)}(\sigma)\Bigr]^{2}\Biggr\}, \\[2.5mm]
\ro{(\ell)}{4}(\sigma) &=& 4 \AIm{(\ell)}(\sigma)
     \ARe{(\ell)}(\sigma)
     \Biggl\{\!\Bigl[\ARe{(\ell)}(\sigma)\Bigr]^{2} -
     \pi^{2}\Bigl[\AIm{(\ell)}(\sigma)\Bigr]^{2}\Biggr\}.\hspace{10mm}
\end{eqnarray}

Then, one has to calculate (by hands) the real and imaginary
parts~(\ref{AReImDef}) of the couplants $a\ind{(\ell)}{s}(Q^2)$
(\ref{aPert1L})--(\ref{aPert4L}) on a physical cut. In turn, this will
enable one to construct the spectral functions
$\ro{(\ell)}{j}(\sigma)$~(\ref{RhoPartHO}) corresponding to any integer
power~($j \ge 1$) of the couplant~$a\ind{(\ell)}{s}(Q^2)$ up to the
four--loop level. The explicit expressions for the calculated
functions~$\AIm{(\ell)}(\sigma)$ and~$\ARe{(\ell)}(\sigma)$ ($1 \le \ell
\le 4$) are given in App.~\ref{Sect:Rho} (see also App.~C of
Ref.~\cite{Prosperi3} and App.~C of Ref.~\cite{Review}). The four--loop
($\ell=4$) spectral functions $\ro{(4)}{j}(\sigma)$ ($1 \le j \le 4$) are
shown in Fig.~\ref{Plot:Rho}.

\section{Basic integrals}
\label{Sect:Int}

The explicit expressions for the spectral functions
$\ro{(\ell)}{j}(\sigma)$~(\ref{RhoPartHO}) enable one to compute the
corresponding nonpower expansion functions in spacelike (Eqs.~(\ref{dAPT})
and~(\ref{ASLMAPT})) and timelike (Eqs.~(\ref{rAPT}) and~(\ref{ATLMAPT}))
domains. Specifically, in the framework of massless APT the spacelike
expansion functions~(\ref{dAPT}) read
\begin{equation}
\label{ASL}
\bar A_{\mbox{\tiny SL}}(z) = \int\limits_{0}^{\infty}
\frac{\varrho(\sigma)}{\sigma + z} \, d \sigma.
\end{equation}
For the numerical evaluation of~$\bar A_{\mbox{\tiny SL}}(z)$ it is
convenient to split Eq.~(\ref{ASL}) into three terms\footnote{This was
first proposed by Dr.~I.L.$\,$Solovtsov.}, namely
\begin{equation}
\label{ASLNum}
\bar A_{\mbox{\tiny SL}}(z) =
\int\limits_{-1}^{0}
\frac{r(x) e^{1/x}}{e^{1/x} + z} \, d x +
\int\limits_{-1}^{1}
\frac{\varrho_{y}(y)}{1 + z e^{-y}} \, d y +
\int\limits_{0}^{1}
\frac{r(x)}{1 + z e^{-1/x}} \, d x,
\end{equation}
where
\begin{equation}
\label{RhoXYDef}
\varrho_{y}(y) = \varrho(\sigma)\biggr|_{\sigma=e^y}, \qquad
\varrho_{x}(x) = \varrho_{y}(y)\biggr|_{y=1/x}, \qquad
r(x) = \frac{\varrho_{x}(x)}{x^2}.
\end{equation}
The numerical integration of Eq.~(\ref{ASLNum}) is implemented within
QCDMAPT library by the subroutine \texttt{APTSL} (see
Sect.~\ref{Sect:Template}). In turn, the timelike massless APT expansion
functions~(\ref{rAPT}) take the form
\begin{equation}
\label{ATL}
\bar A_{\mbox{\tiny TL}}(w) = \int\limits_{w}^{\infty}
\varrho(\sigma)\,\frac{d \sigma}{\sigma}.
\end{equation}
Similarly to the previous case, it is worth representing this equation in
the following way:
\begin{figure}[t]
\centerline{\includegraphics[width=110mm,clip]{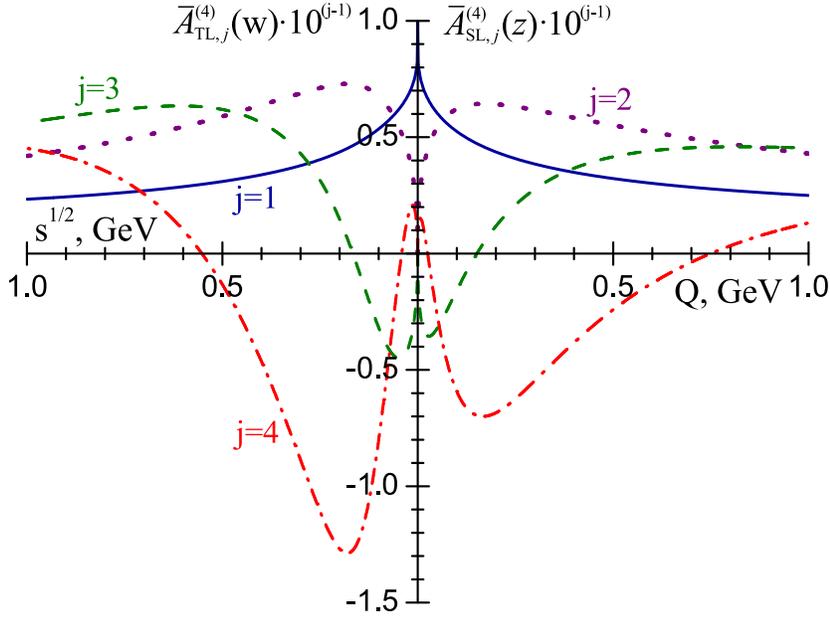}}
\caption{The four--loop ``massless'' APT expansion functions
$\bar\ASL{(4)}{j}(z)$~(\ref{dAPT}) and $\bar\ATL{(4)}{j}(w)$~(\ref{rAPT}):
$j=1$~(solid curves), $j=2$~(dotted curves, scaled$\,\times 10$),
$j=3$ (dashed curves, scaled$\,\times 10^2$), and $j=4$~(dot--dashed
curves, scaled$\,\times 10^3$). The values of parameters: $\nf=3$,
$\Lambda=350\,$MeV, $z=Q^2/\Lambda^2$, $w=s/\Lambda^2$.}
\label{Plot:APT}
\end{figure}
\begin{equation}
\label{ATLNum}
\bar A_{\mbox{\tiny TL}}(w) = \left\{
\begin{array}{l}
J_1\biggl(-1, \displaystyle{\frac{1}{\ln w}}\biggr) +
J_2(-1, 1) + J_1(0, 1), \qquad \ln w < -1 \\[4.5mm]
J_2(\ln w, 1) + J_1(0, 1), \qquad -1 \le \ln w < 1 \\[2.5mm]
J_1\biggl(0, \displaystyle{\frac{1}{\ln w}}\biggr), \qquad 1 \le \ln w \\
\end{array}
\right.
\end{equation}
where
\begin{equation}
\label{ATLIntAux}
J_1(x_1,x_2) = \int\limits_{x_1}^{x_2} r(x)\, d x, \qquad \quad
J_2(y_1,y_2) = \int\limits_{y_1}^{y_2} \varrho_{y}(y)\, d y.
\end{equation}
The numerical integration of Eq.~(\ref{ATLNum}) is implemented within
QCDMAPT library by the subroutine \texttt{APTTL} (see
Sect.~\ref{Sect:Template}). The four--loop massless APT expansion
functions~$\bar\ASL{(4)}{j}(z)$ and~$\bar\ATL{(4)}{j}(w)$ ($1 \le j \le
4$) are shown in Fig.~\ref{Plot:APT}.

\begin{figure}[t]
\centerline{\includegraphics[width=110mm,clip]{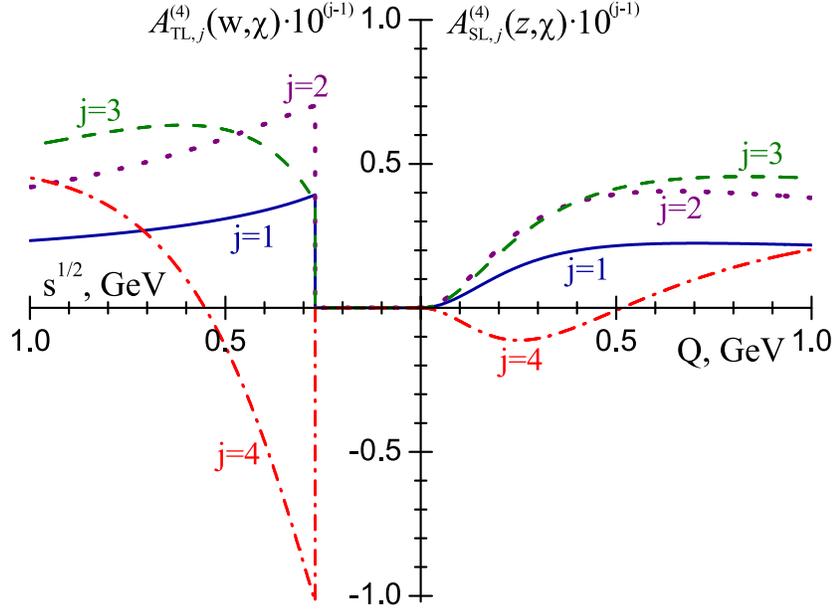}}
\caption{The four--loop MAPT expansion functions
$\ASL{(4)}{j}(z,\chi)$~(\ref{ASLMAPT}) and
$\ATL{(4)}{j}(w,\chi)$~(\ref{ATLMAPT}): $j=1$~(solid curves),
$j=2$~(dotted curves, scaled$\,\times 10$), $j=3$~(dashed curves,
scaled$\,\times 10^2$), and $j=4$~(dot--dashed curves, scaled$\,\times
10^3$). The values of parameters: $\nf=3$, $\Lambda=350\,$MeV,
$m=270\,$MeV, $z=Q^2/\Lambda^2$, $w=s/\Lambda^2$.}
\label{Plot:MAPT}
\end{figure}

In the framework of massive APT the spacelike expansion
functions~(\ref{ASLMAPT}) read
\begin{equation}
\label{ASLM}
A_{\mbox{\tiny SL}}(z,\chi) = \frac{z}{z+\chi}
\int\limits_{\chi}^{\infty} \varrho(\sigma)\,
\frac{\sigma - \chi}{\sigma+z}\, \frac{d \sigma}{\sigma}.
\end{equation}
For the numerical evaluation of~$A_{\mbox{\tiny SL}}(z,\chi)$ it is
convenient to represent Eq.~(\ref{ASLM}) in the following way:
\begin{equation}
\label{ASLMNum}
A_{\mbox{\tiny SL}}(z,\chi)  = \left\{
\begin{array}{l}
\!{\displaystyle\frac{z}{z+\chi}}
\biggl[I_1\biggl(-1, \displaystyle{\frac{1}{\ln\chi}}\biggr) +
\!I_2(-1, 1) + I_3(0, 1)\biggr],
\qquad  \ln\chi < -1 \\[7mm]
\!{\displaystyle\frac{z}{z+\chi}}
\Bigl[I_2(\ln\chi, 1) + I_3(0, 1)\Bigr],
\qquad -1 \le \ln\chi < 1 \\[5mm]
\!{\displaystyle\frac{z}{z+\chi}}\,
I_3\biggl(0, \displaystyle{\frac{1}{\ln\chi}}\biggr), \qquad 1 \le \ln\chi \\
\end{array}
\right.
\end{equation}
where
\begin{equation}
I_{1}(x_1,x_2) = \int\limits_{x_1}^{x_2} r(x)\,
\frac{e^{1/x}-\chi}{e^{1/x}+z}\, d x,
\end{equation}
\begin{equation}
I_{2}(y_1,y_2) = \int\limits_{y_1}^{y_2} \varrho_{y}(y)\,
\frac{e^{y}-\chi}{e^{y}+z}\, d y,
\end{equation}
\begin{equation}
\label{ASLMIntAux}
I_{3}(x_1,x_2) = \int\limits_{x_1}^{x_2} r(x)\,
\frac{1-\chi e^{-1/x}}{1+z e^{-1/x}}\, d x.
\end{equation}
The numerical integration of Eq.~(\ref{ASLMNum}) is implemented within
QCDMAPT library by the subroutine \texttt{MAPTSL} (see
Sect.~\ref{Sect:Template}). In turn, the timelike MAPT expansion
functions~(\ref{ATLMAPT}) take the form
\begin{equation}
\label{ATLM}
A_{\mbox{\tiny TL}}(w,\chi) = \theta(w-\chi)\int\limits_{w}^{\infty}
\varrho(\sigma)\,\frac{d \sigma}{\sigma} =
\theta(w-\chi) \bar A_{\mbox{\tiny TL}}(w),
\end{equation}
where $\bar A_{\mbox{\tiny TL}}(w)$ is defined in Eq.~(\ref{ATLNum}). The
numerical integration of Eq.~(\ref{ATLM}) is implemented within QCDMAPT
library by the subroutine \texttt{MAPTTL} (see Sect.~\ref{Sect:Template}).
The four--loop MAPT expansion functions~$\ASL{(4)}{j}(z,\chi)$
and~$\ATL{(4)}{j}(w,\chi)$ ($1 \le j \le 4$) are shown in
Fig.~\ref{Plot:MAPT}.

\section{Description of the template program}
\label{Sect:Template}

The provided template program~(\texttt{template.mw}) is self--explaining.
Besides, both template program and QCDMAPT library (\texttt{QCDMAPT.lib})
contain detailed comments.

The template program is organized as follows (see~\texttt{template.pdf}).
On the first page the QCDMAPT library is loaded and a sample set of input
parameters is specified: $\nf=3$~active quarks, $Q=700\,$MeV,
$s=(700\,\mbox{MeV})^2$, $m=270\,$MeV, and $\Lambda=350\,$MeV. The
evaluation of expansion functions for the given set of parameters is
presented on page~2. In particular, the first computation deals with the
first--order~($j=1$) $\ell$--loop ($\ell=1,2,3,4$) expansion functions,
namely, $a\ind{(\ell)}{s}(Q^2)$ (Eqs.~(\ref{aPert1L})--(\ref{aPert4L})),
$\bar\ASL{(\ell)}{1}(z)$ (Eq.~(\ref{dAPT})),
$\bar\ATL{(\ell)}{1}(w)$ (Eq.~(\ref{rAPT})),
$\ASL{(\ell)}{1}(z,\chi)$ (Eq.~(\ref{ASLMAPT})), and
$\ATL{(\ell)}{1}(w,\chi)$ (Eq.~(\ref{ATLMAPT})). The second
computation evaluates four--loop ($\ell=4$) expansion functions of $j$--th
order ($j=1,2,3,4$), namely,
$\bigl[a\ind{(4)}{s}(Q^2)\bigr]^{j}$ (Eq.~(\ref{aPert4L})),
$\bar\ASL{(4)}{j}(z)$ (Eq.~(\ref{dAPT})),
$\bar\ATL{(4)}{j}(w)$ (Eq.~(\ref{rAPT})),
$\ASL{(4)}{j}(z,\chi)$ (Eq.~(\ref{ASLMAPT})),
and $\ATL{(4)}{j}(w,\chi)$ (Eq.~(\ref{ATLMAPT})). The data arrays of the first--order one--loop
expansion functions are computed and relevant plots are presented on
page~3 (see Fig.~\ref{Plot:A1L} and its caption for the details). The data
arrays of the massless APT four--loop ($\ell=4$) expansion functions of
$j$--th order ($j=1,2,3,4$) are computed and corresponding plots are
presented on page~4 (see Fig.~\ref{Plot:APT} and its caption for the
details). The data arrays of the MAPT four--loop ($\ell=4$) expansion
functions of $j$--th order ($j=1,2,3,4$) are computed and relevant plots
are presented on page~5 (see Fig.~\ref{Plot:MAPT} and its caption for the
details).

The summary of QCDMAPT package commands is given below.

\newpage

\subheader{PERTURBATIVE QCD}

\begin{description}

\item
\cmd{AlphaPert$\,\ell\,$L(z)}$=\al{(\ell)}{s}(Q^2)$ ($\ell=1,2,3,4$;
$z=Q^2/\Lambda^2$): $\ell$--loop perturbative strong running coupling,
see App.~\ref{Sect:RCPert}.

\vskip2.5mm

\item
\cmd{APert$\,\ell\,$L(z)}$=a\ind{(\ell)}{s}(Q^2) =
\al{(\ell)}{s}(Q^2)\beta_{0}/(4\pi)$ ($\ell=1,2,3,4$; $z=Q^2/\Lambda^2$):
$\ell$--loop perturbative QCD couplant, see
Eqs.~(\ref{aPert1L})--(\ref{aPert4L}).

\vskip2.5mm

\item
\cmd{beta$\,j$}$= \beta_{j}$ ($j=0,1,2,3$): perturbative
($j+1$)--loop expansion coefficient of the QCD $\beta$~function,
see Eqs.~(\ref{BetaPert1L})--(\ref{BetaPert4L}).

\vskip2.5mm

\item
\cmd{B$\,j$}$= B_{j} = \beta_{j}/\beta_{0}^{j+1}$ ($j=1,2,3$): combination
of QCD $\beta$~function perturbative expansion coefficients, see
App.~\ref{Sect:RCPert}.

\end{description}

\subheader{SPECTRAL FUNCTIONS}

\begin{description}

\item
\cmd{ARe$\,\ell\,$LY(y)}$=\ARe{(\ell)}(\sigma)\Bigr|_{\sigma=e^y}$
($\ell=1,2,3,4$): real part of the $\ell$--loop perturbative QCD couplant
on a physical cut, see Eqs.~(\ref{AReImDef}), (\ref{RhoXYDef}), and
App.~\ref{Sect:Rho}.

\vskip2.5mm

\item
\cmd{AIm$\,\ell\,$LY(y)}$=\AIm{(\ell)}(\sigma)\Bigr|_{\sigma=e^y}$
($\ell=1,2,3,4$): imaginary part of the $\ell$--loop perturbative QCD
couplant on a physical cut, see Eqs.~(\ref{AReImDef}), (\ref{RhoXYDef}),
and App.~\ref{Sect:Rho}.

\vskip2.5mm

\item
\cmd{Rho$\,\ell\,$L$\,j\,$pY(y)}$=\ro{(\ell)}{j}(\sigma)\Bigr|_{\sigma=e^y}$
($\ell=1,2,3,4$; $j = 1,...\,,\ell$): spectral function (in terms of
$y=\ln\sigma$ variable) corresponding to $j$--th power of the $\ell$--loop
perturbative QCD couplant, see Eqs.~(\ref{RhoDef}), (\ref{RhoPartHO}),
(\ref{RhoXYDef}), and App.~\ref{Sect:Rho}.

\vskip2.5mm

\item
\cmd{R$\,\ell\,$L$\,j\,$pX(x)}$= r_{j}^{(\ell)}(x) =
x^{-2}\ro{(\ell)}{j}(\sigma)\Bigr|_{\sigma=e^{1/x}}$ ($\ell=1,2,3,4$; $j =
1,...\,,\ell$): spectral function (in terms of $x=1/\ln\sigma$ variable)
corresponding to $j$--th power of the $\ell$--loop perturbative QCD
couplant, see Eqs.~(\ref{RhoDef}), (\ref{RhoPartHO}), (\ref{RhoXYDef}),
and App.~\ref{Sect:Rho}.

\end{description}

\subheader{BASIC INTEGRALS}

\begin{description}

\item
\cmd{APTSL(RhoY,RX,z)}$=\bar A_{\mbox{\tiny SL}}(z)$: numerical
integration for the massless APT spacelike expansion functions, see
Eqs.~(\ref{ASL})--(\ref{RhoXYDef}) with $\ro{}{y}(y)=\,$\texttt{RhoY(y)}
and $r(x)=\,$\texttt{RX(x)}.

\vskip2.5mm

\item
\cmd{APTTL(RhoY,RX,w)}$=\bar A_{\mbox{\tiny TL}}(w)$: numerical
integration for the massless APT timelike expansion functions, see
Eqs.~(\ref{ATL})--(\ref{ATLIntAux}), (\ref{RhoXYDef}) with
$\ro{}{y}(y)=\,$\texttt{RhoY(y)} and $r(x)=\,$\texttt{RX(x)}.

\vskip2.5mm

\item
\cmd{MAPTSL(RhoY,RX,z,chi)}$=A_{\mbox{\tiny SL}}(z,\chi)$: numerical
integration for the MAPT spacelike expansion functions, see
Eqs.~(\ref{ASLM})--(\ref{ASLMIntAux}), (\ref{RhoXYDef}) with
$\ro{}{y}(y)=\,$\texttt{RhoY(y)} and $r(x)=\,$\texttt{RX(x)}.

\vskip2.5mm

\item
\cmd{MAPTTL(RhoY,RX,w,chi)}$=A_{\mbox{\tiny TL}}(w,\chi)$: numerical
integration for the MAPT timelike expansion functions, see
Eqs.~(\ref{ATLM}), (\ref{ATL})--(\ref{ATLIntAux}), (\ref{RhoXYDef}) with
$\ro{}{y}(y)=\,$\texttt{RhoY(y)} and $r(x)=\,$\texttt{RX(x)}.

\end{description}

\subheader{APT EXPANSION FUNCTIONS}

\begin{description}

\item
\cmd{ASL$\,\ell\,$L$\,j\,$p(z)}$= \bar\ASL{(\ell)}{j}(z)$ ($\ell =
1,2,3,4$; $j = 1,...\,,\ell$): APT spacelike expansion function
($\ell$--loop level, $j$--th order), see Eq.~(\ref{dAPT}).

\vskip2.5mm

\item
\cmd{ATL$\,\ell\,$L$\,j\,$p(w)}$= \bar\ATL{(\ell)}{j}(w)$ ($\ell =
1,2,3,4$; $j = 1,...\,,\ell$): APT timelike expansion function
($\ell$--loop level, $j$--th order), see Eq.~(\ref{rAPT}).

\vskip2.5mm

\item
\cmd{AlphaSL$\,\ell\,$L(z)}$= \bar\ASL{(\ell)}{1}(z)\, 4\pi/\beta_{0}$
($\ell = 1,2,3,4$): APT spacelike \mbox{$\ell$--loop} effective
coupling, see Eq.~(\ref{dAPT}).

\vskip2.5mm

\item
\cmd{AlphaTL$\,\ell\,$L(w)}$= \bar\ATL{(\ell)}{1}(w)\, 4\pi/\beta_{0}$
($\ell = 1,2,3,4$): APT timelike \mbox{$\ell$--loop} effective
coupling, see Eq.~(\ref{rAPT}).

\end{description}

\subheader{MAPT EXPANSION FUNCTIONS}

\begin{description}

\item
\cmd{ASLm$\,\ell\,$L$\,j\,$p(z,chi)}$= \ASL{(\ell)}{j}(z,\chi)$
($\ell = 1,2,3,4$; $j = 1,...\,,\ell$): MAPT spacelike expansion function
($\ell$--loop level, $j$--th order), see Eq.~(\ref{ASLMAPT}).

\vskip2.5mm

\item
\cmd{ATLm$\,\ell\,$L$\,j\,$p(w,chi)}$= \ATL{(\ell)}{j}(w,\chi)$
($\ell = 1,2,3,4$; $j = 1,...\,,\ell$): MAPT timelike expansion function
($\ell$--loop level, $j$--th order), see Eq.~(\ref{ATLMAPT}).

\vskip2.5mm

\item
\cmd{AlphaSLm$\,\ell\,$L(z,chi)}$= \ASL{(\ell)}{1}(z,\chi)\,
4\pi/\beta_{0}$ ($\ell = 1,2,3,4$): MAPT spacelike \mbox{$\ell$--loop}
effective coupling, see Eq.~(\ref{ASLMAPT}).

\vskip2.5mm

\item
\cmd{AlphaTLm$\,\ell\,$L(w,chi)}$= \ATL{(\ell)}{1}(w,\chi)\,
4\pi/\beta_{0}$ ($\ell = 1,2,3,4$): MAPT timelike \mbox{$\ell$--loop}
effective coupling, see Eq.~(\ref{ATLMAPT}).

\end{description}

\section{Conclusions}
\label{Sect:Concl}

A program package, which facilitates computations in the framework of
the Analytic approach to QCD, is developed. It includes the explicit
expressions for relevant spectral functions calculated up to the
four--loop level and the subroutines for necessary integrals.

\vskip5mm

\noindent
This work was partially supported by the grants RFBR-08-01-00686,
BRFBR-JINR-F08D-001, NS-1027.2008.2, and JINR grant.

\appendix

\section{Perturbative QCD running coupling}
\label{Sect:RCPert}

This Section contains explicit expressions for the perturbative strong
running coupling $\al{(\ell)}{s}(Q^2) \equiv 4\pi\,
a\ind{(\ell)}{s}(Q^2)/\beta_{0}$ up to the four--loop level. Specifically,
\begin{eqnarray}
\label{aPert1L}
a\ind{(1)}{s}(Q^2) &=& \frac{1}{\ln z}, \qquad z=\frac{Q^2}{\Lambda^2}, \\[2.5mm]
\label{aPert2L}
a\ind{(2)}{s}(Q^2) &=& \frac{1}{\ln z}-B_{1}\frac{\ln(\ln z)}{\ln^{2} z}, \\[2.5mm]
\label{aPert3L}
a\ind{(3)}{s}(Q^2) &=& \frac{1}{\ln z}-B_{1}\frac{\ln(\ln z)}{\ln^{2} z}
\nonumber \\[2.5mm] &&
+\frac{1}{\ln^{3}\!z}\biggl\{B_{1}^{2} \Bigl[\ln^2(\ln z)
- \ln(\ln z)-1\Bigr]+B_{2}\biggr\}, \\[2.5mm]
\label{aPert4L}
a\ind{(4)}{s}(Q^2) &=& \frac{1}{\ln z} - B_{1} \frac{\ln(\ln z)}{\ln^{2} z}
\nonumber \\[2.5mm] &&
+\frac{1}{\ln^{3}\!z}\biggl\{B_{1}^{2} \Bigl[\ln^2(\ln z)
-\ln(\ln z)-1\Bigr]+B_{2}\biggr\}
\nonumber \\[2.5mm] &&
+\frac{1}{\ln^{4}\!z}\Biggl\{B_{1}^{3} \biggl[-\ln^{3}(\ln z)
+\frac{5}{2}\ln^{2}(\ln z)
\nonumber \\[2.5mm] &&
+ 2 \ln(\ln z) -\frac{1}{2}\biggr]
- 3B_{1}B_{2}\ln(\ln z) + \frac{1}{2}B_{3}\Biggr\},\hspace{5mm}
\end{eqnarray}
where $B_j=\beta_{j}/\beta_{0}^{j+1}$ is the combination of QCD
$\beta$~function perturbative expansion coefficients:
\begin{eqnarray}
\label{BetaPert1L}
\beta_{0} &=& 11 - \frac{2}{3}\nf, \\[2.5mm]
\label{BetaPert2L}
\beta_{1} &=& 102 - \frac{38}{3}\nf, \\[2.5mm]
\label{BetaPert3L}
\beta_{2} &=& \frac{2857}{2} - \frac{5033}{18}\nf +
              \frac{325}{54}\nf^2, \\[2.5mm]
\label{BetaPert4L}
\beta_{3} &=& \frac{149753}{6} + 3564\zeta(3) -
\left[\frac{1078361}{162}+\frac{6508}{27}\zeta(3)\right]\!\nf
\nonumber \\[2.5mm] &&
+ \left[\frac{50065}{162}+\frac{6472}{81}\zeta(3)\right]\!\nf^{2} +
\frac{1093}{729}\nf^{3}.
\end{eqnarray}
In these equations $\nf$ stands for the number of active quarks and
$\zeta(x)$ denotes the Riemann $\zeta$~function, $\zeta(3) \simeq 1.202$.
The one-- and two--loop coefficients ($\beta_{0}$~and~$\beta_{1}$) are
scheme--independent, whereas the expressions given for~$\beta_{2}$
and~$\beta_{3}$ are calculated in the \MSbar~subtraction scheme (see
papers~\cite{BetaPert1L,BetaPert2L,BetaPert3L,BetaPert4L} and references
therein for the details). Note that expressions
(\ref{aPert2L})--(\ref{aPert4L}) correspond to approximate solutions of
the perturbative renormalization group~(RG) equation for the QCD invariant
charge, see, e.g., Ref.~\cite{Lambert}, Sect.~2 of
review~\cite{Prosperi4}, and App.~A of review~\cite{Review}. Nonetheless,
the difference between the approximate running
coupling~$\al{(\ell)}{s}(Q^2)$ (\ref{aPert2L})--(\ref{aPert4L}) and the
exact solution of perturbative RG~equation~$\al{(\ell)}{ex}(Q^2)$ is not
controllable at every considered loop level.

\section{Spectral functions}
\label{Sect:Rho}

As it has been mentioned in Sect.~\ref{Sect:RhoCalc}, the spectral
functions $\ro{(\ell)}{j}(\sigma)$~(\ref{RhoDef}) can be represented
in the following form:
\begin{equation}
\ro{(\ell)}{j}(\sigma) = \sum_{k=0}^{K_j}
{j \choose 2k+1}
(-1)^{k}\, \pi^{2k}
\Bigl[\AIm{(\ell)}(\sigma)\Bigr]^{2k+1}
\Bigl[\ARe{(\ell)}(\sigma)\Bigr]^{j-2k-1},
\end{equation}
where
\begin{equation}
\lim_{\varepsilon \to 0_{+}}
a\ind{(\ell)}{s}\Bigl[-\Lambda^2(\sigma \mp i\varepsilon)\Bigr] \equiv
\ARe{(\ell)}(\sigma) \mp i \pi \AIm{(\ell)}(\sigma),
\end{equation}
$\ell$--loop perturbative couplants $a\ind{(\ell)}{s}(Q^2)$ are given by
Eqs.~(\ref{aPert1L})--(\ref{aPert4L}), and $K_j = j/2 + (j \;\mbox{mod}\;
2)/2 -1$. The functions $\AIm{(\ell)}(\sigma)$ and $\ARe{(\ell)}(\sigma)$,
calculated up to the four--loop level ($1 \le \ell \le 4$), are listed
below.

\vskip5mm

\noindent
One--loop level:
\nopagebreak
\begin{equation}
\AIm{(1)}(\sigma) = \frac{1}{y^2+\pi^2}, \qquad\quad
\ARe{(1)}(\sigma) = \frac{y}{y^2+\pi^2}, \qquad\quad
y=\ln\sigma.
\end{equation}

\vskip5mm

\noindent
Two--loop level:
\nopagebreak
\begin{eqnarray}
\AIm{(2)}(\sigma) &=& \frac{1}{(y^2+\pi^2)^{2}}
     \Bigl[2 y \Sy{2}{1} + (\pi^2-y^2)\Sy{2}{2}\Bigr], \\[1.5mm]
\ARe{(2)}(\sigma) &=& \frac{1}{(y^2+\pi^2)^{2}}
     \Bigl[(y^2-\pi^2)\Sy{2}{1} + 2 \pi^2 y \Sy{2}{2}\Bigr],
\end{eqnarray}
where
\begin{eqnarray}
\Sy{2}{1} &=& y - \frac{B_{1}}{2} \ln(y^{2}+\pi^{2}), \\[1.5mm]
\Sy{2}{2} &=& 1 - B_{1} \biggl[\frac{1}{2}-\frac{1}{\pi}
     \arctan\!\left(\frac{y}{\pi}\right)\biggr]\!,
\end{eqnarray}
and $B_j=\beta_{j}/\beta_{0}^{j+1}$ is the combination of QCD
$\beta$~function perturbative expansion coefficients (see
App.~\ref{Sect:RCPert}).

\noindent
Three--loop level:
\nopagebreak
\begin{eqnarray}
\AIm{(3)}(\sigma) &=& \frac{1}{(y^{2}+\pi^{2})^{3}}\Bigl[
     (3y^{2}-\pi^{2})\Sy{3}{3}
- y(y^{2}-3\pi^{2})\Sy{3}{4}\Bigr]\!, \\[1.5mm]
\ARe{(3)}(\sigma) &=& \frac{1}{(y^{2}+\pi^{2})^{3}}\Bigl[
     y(y^{2}-3\pi^{2})\Sy{3}{3}
+ \pi^2 (3y^{2}-\pi^{2})\Sy{3}{4}\Bigr]\!, \hspace{15mm}
\end{eqnarray}
where
\begin{eqnarray}
\Sy{3}{1} &=& \frac{1}{2} \ln(y^{2}+\pi^{2}), \\[1.5mm]
\Sy{3}{2} &=& \frac{1}{2}-\frac{1}{\pi}
     \arctan\!\left(\frac{y}{\pi}\right), \\[1.5mm]
\Sy{3}{3} &=& y^2-\pi^2-B_1\!\left[y\Sy{3}{1}-\pi^2\Sy{3}{2}\right] \nonumber \\[1.5mm]
          && + B_{1}^{2} \biggl\{\!\Sy{3}{1}\!\left[\Sy{3}{1}-1\right]
           - \pi^{2}\!\left[\Sy{3}{2}\right]^{2}-1\!\biggr\} + B_{2}, \\[1.5mm]
\Sy{3}{4} &=& 2y - B_{1}\!\left[\Sy{3}{1}+y\Sy{3}{2}\right]\!
          + B_{1}^{2}\Sy{3}{2}\!\left[2\Sy{3}{1}-1\right]. \hspace{15mm}
\end{eqnarray}

\vskip5mm

\noindent
Four--loop level:
\nopagebreak
\begin{eqnarray}
\AIm{(4)}(\sigma) &=& \frac{1}{(y^{2}+\pi^{2})^{4}}\biggl\{\!4y(y^2-\pi^2)\!
     \left[\Sy{4}{3}+\Sy{4}{5}+\Sy{4}{7}\right] \nonumber \\[1.5mm]
     &&+ \Bigl[4\pi^2y^2-(y^2-\pi^2)^2\Bigr]\!
     \Bigl[\Sy{4}{4}+\Sy{4}{6}+\Sy{4}{8}\Bigr]\!
     \biggr\}\!, \\[1.5mm]
\ARe{(4)}(\sigma) &=& \frac{1}{(y^{2}+\pi^{2})^{4}}\biggl\{\!
     \Bigl[(y^2-\pi^2)^{2}-4\pi^2y^2\Bigr]\!
     \Bigl[\Sy{4}{3}+\Sy{4}{5}+\Sy{4}{7}\Bigr] \nonumber \\[1.5mm]
     &&+ 4\pi^2y(y^2-\pi^2)\!
     \left[\Sy{4}{4}+\Sy{4}{6}+\Sy{4}{8}\right]\!
     \biggr\}\!,\hspace{5mm}
\end{eqnarray}
where
\begin{eqnarray}
\Sy{4}{1} &=& \frac{1}{2} \ln(y^{2}+\pi^{2}), \\[0.5mm]
\Sy{4}{2} &=& \frac{1}{2}-\frac{1}{\pi}
     \arctan\!\left(\frac{y}{\pi}\right)\!, \\[0.5mm]
\Sy{4}{3} &=& B_{1}^{2}\,y\biggl\{\!\Sy{4}{1}\!\left[\Sy{4}{1}-1\right]
     - \pi^2 \!\left[\Sy{4}{2}\right]^{2} +
     \frac{B_2}{B_{1}^{2}}-1\!\biggr\} \hspace{10mm}\nonumber \\[0.5mm]
     &&- B_{1}^{2}\,\pi^2\Sy{4}{2}\!
     \left[2\Sy{4}{1}-1\right]\!, \\[0.5mm]
\Sy{4}{4} &=& B_{1}^{2} \biggl\{\!\Sy{4}{1}\!\left[\Sy{4}{1}-1\right]-
     \pi^2\! \left[\Sy{4}{2}\right]^{2} \nonumber \\[0.5mm]
     &&+\, y\,\Sy{4}{2}\!\left[2\Sy{4}{1}-1\right] +
     \frac{B_2}{B_{1}^{2}}-1\biggr\}\!, \\[1.5mm]
\Sy{4}{5} &=& B_{1}^{3}
     \Biggl\{\!\Sy{4}{1}\!\left[3\pi^2\!\left(\Sy{4}{2}\right)^{2}
     - \left(\Sy{4}{1}\right)^{2}\right] \nonumber \\[1.5mm]
     &&+\, \frac{5}{2}\left[\left(\Sy{4}{1}\right)^{2}
     - \pi^2\left(\Sy{4}{2}\right)^{2}\right] \nonumber \\[1.5mm]
     &&+\, \Sy{4}{1}\!\left(2-3\,\frac{B_2}{B_{1}^{2}}\right)
     + \frac{1}{2}\!\left(\frac{B_3}{B_{1}^{3}}-1\right)\!\Biggr\}\!, \\[1.5mm]
\Sy{4}{6} &=& B_{1}^{3} \Sy{4}{2} \biggl\{\!\pi^2\!\left[\Sy{4}{2}\right]^{2}
     - 3\!\left[\Sy{4}{1}\right]^{2} \nonumber \\[1.5mm]
     &&+\, 5 \Sy{4}{1} - 3\frac{B_2}{B_{1}^{2}}+2\biggr\}\!, \\[1.5mm]
\Sy{4}{7} &=& y(y^{2}-3\pi^{2})
     + B_{1} \Bigl[ 2\pi^{2}y\Sy{4}{2}
     - (y^{2}-\pi^{2})\Sy{4}{1}\Bigr]\!, \hspace{10mm} \\[1.5mm]
\Sy{4}{8} &=& 3y^{2}-\pi^{2} - B_{1} \Bigl[(y^{2}-\pi^{2})\Sy{4}{2} +
     2y\Sy{4}{1}\Bigr]\!.
\end{eqnarray}

\end{document}